\documentclass{article}[12]
\usepackage{graphicx,amssymb,amsfonts,bm,mathptmx,enumerate,eucal}
\usepackage[polish,english]{babel}
\usepackage[tbtags]{amsmath}
\begin{document}
\title{Michelson--Morley experiment revisited\footnote{On the occasion of the {\it World Year of Physics 2oo5}.}}
\selectlanguage{polish}

\author{Bogus"law Broda and Marcin Ostrowski}
\date{}
\maketitle

\noindent
{\it Department of Theoretical Physics, University of Lodz,}\\
{\it ul.~Pomorska 149/153, 90-236 {\L}{\'o}d{\'z}, Poland}\\
{\it bobroda@uni.lodz.pl},
{\it m.ostrowski@merlin.fic.uni.lodz.pl}

\selectlanguage{english}
\begin{abstract}
The idea of the Michelson--Morley experiment is theoretically reanalyzed. Elementary arguments are put forward to precisely derive the most general allowable form of the directional dependence of the one-way velocity of light.
\end{abstract}

According to XIX-century physics light was supposed to propagate in the aether, a mysterious medium devised especially for this
purpose. As light was to travel with respect to the aether with a fixed velocity, an experiment was suggested to detect the
dependence of the velocity on the direction in a moving frame of the laboratory on the Earth. The experiment was proposed
and the first time performed by Albert Michelson (a Nobel laureate, American physicist born in Poland) \cite{raz}. The
experiment is continually being repeated, known as the Michelson--Morley (MM) experiment, with ever increasing accuracy
and improved technical realization (see, e.g. \cite{dwa}).
The result is always the same, negative, i.e. no dependence of the velocity of light on the direction has been ever detected, at least, this is a generally accepted (but sometimes disputable \cite{Consoli}) conclusion. As a (standard) consequence, the velocity of light is the same in each inertial frame, in any direction, and no aether exists.

Strictly speaking, the experimental results are usually translated into the theoretical statement saying that the {\it average} velocity of light around any closed path is constant and equal to the universal constant $c$. According to \cite{Edwards} (see also Chapt.~1.3 in \cite{cztery}) this bound yields the following directional (angular) dependence of the one-way velocity of light
\begin{equation}
\label{First}
c(\theta)={c\over1-\lambda\cos\theta},
\end{equation}
where $\theta$ is the angle between the direction of light and ``the direction of anisotropy'', and $\lambda$ is a parameter belonging to the interval from $0$ to $1$.

An elementary theoretical analysis of the actual MM experiments, proposed in this paper, yields a more general solution. Among other things, we show that
the negative result of the MM experiments imposes milder
constrains on the directional dependence (anisotropy) of the one-way velocity of light. In particular, the statement that the average velocity of light along any closed path is constant is too strong from our point of view because it does not strictly follows from actual MM experiments.

Let us briefly recall the idea of the MM experiment. In any, traditional or modern, version of the experiment we compare the
differences of the time of the travel of light in two orthogonal directions (vertical and horizontal, say) in two positions
(primary and final) differing by $90^\circ$.
To arrive at our result we assume full generality. Therefore, let us introduce the following notation (see Fig.~1):

$c_+^\perp$ --- forward vertical velocity of light,

$c_-^\perp$ --- backward vertical velocity of light,

$c_+^\shortparallel$ --- forward horizontal velocity of light,

$c_-^\shortparallel$ --- backward horizontal velocity of light,

\noindent
and

$L_1$ --- primarily vertical route,

$L_2$ --- primarily horizontal route.

\begin{figure}
\begin{center}
\includegraphics[width=5cm]{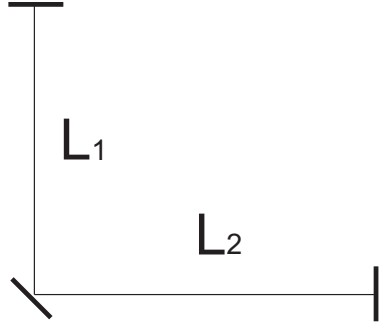}
\end{center}
\caption{The Michelson--Morley interferometer.}
\end{figure}
\noindent
In primary position, the vertical travel of light takes
\begin{equation}
t_1=\frac{L_1}{c_+^\perp}+\frac{L_1}{c_-^\perp},
\end{equation}
whereas in horizontal direction
\begin{equation}
t_2=\frac{L_2}{c_+^\shortparallel}+\frac{L_2}{c_-^\shortparallel}.
\end{equation}
The difference is
\begin{equation}
\Delta{t}=t_2-t_1=L_2\biggl(\frac{1}{c_+^\shortparallel}+\frac{1}{c_-^\shortparallel}\biggr)-
L_1\biggl(\frac{1}{c_+^\perp}+\frac{1}{c_-^\perp}\biggr).
\end{equation}
After rotation, we have
\begin{equation}
\Delta{t'}=L_2\biggl(\frac{1}{c_+^\perp}+\frac{1}{c_-^\perp}\biggr)-
L_1\biggl(\frac{1}{c_+^\shortparallel}+\frac{1}{c_-^\shortparallel}\biggr).
\end{equation}
The change one could possibly observe is of the form
\begin{equation}
\label{eq:plus}
\Delta{t^*}=\Delta{t'}-\Delta{t}=(L_1+L_2)\biggl[\biggl(\frac{1}{c_+^\perp}+\frac{1}{c_-^\perp}\biggr)-
\biggl(\frac{1}{c_+^\shortparallel}+\frac{1}{c_-^\shortparallel}\biggr)\biggr].
\end{equation}

The negative result of the MM experiment formally means $\Delta{t^*}=0$. (Our analysis is purely theoretical,
and we are not going to engage into the debate whether the equality $\Delta{t^*}=0$ is experimentally well-established or
not, see, e.g.\ \cite{Consoli}.) A very simplified argumentation would say that the equality $\Delta{t^*}=0$ implies
$c_+^\perp=c_-^\perp=c_+^\shortparallel=c_-^\shortparallel=c$.
A bit less simplified argumentation presumes a strictly defined form
of $c_\pm^\perp$, $c_\pm^\shortparallel$ following from geometrical analysis (the Pythagorean theorem and geometrical
addition of velocities) of the movement of inertial systems with respect to the aether, i.e.
\begin{equation}
\label{eq:gwiazdka}
c_\pm^\perp=\sqrt{c^2-v^2}
\end{equation}
and
\begin{equation}
\label{eq:gwiazdka2}
c_\pm^\shortparallel=c\mp{v},
\end{equation}
where $c$ --- velocity of light with respect to the aether, $v$ --- velocity of the inertial system with respect to the aether.
The possibility \eqref{eq:gwiazdka}-\eqref{eq:gwiazdka2} is also excluded by virtue of the experimental fact \mbox{$\Delta{t^*}=0$},
where $\Delta{t^*}$ is defined by Eq.~\eqref{eq:plus}. But there are infinitely many other possibilities consistent with $\Delta{t^*}$
defined by Eq.~\eqref{eq:plus} equal to zero (a particular subset of the possibilities is given by Eq.~\eqref{First}). The aim of the paper is to quantify this fact.

We can rewrite the equation $\Delta{t^*}=0$ with $\Delta{t^*}$ defined by Eq.~\eqref{eq:plus} in the following form
\begin{equation}
\label{ro10}
z_+^\perp+z_-^\perp=z_+^\shortparallel+z_-^\shortparallel,
\end{equation}
where for simplicity we use inverses of the corresponding velocities, $z_{+,-}^{\perp,\shortparallel}\equiv 1/c_{+,-}^{\perp,\shortparallel}$. Eq.~\eqref{ro10} is a functional equation with continuum of
solutions.

\begin{figure}
\begin{center}
\includegraphics[width=6cm]{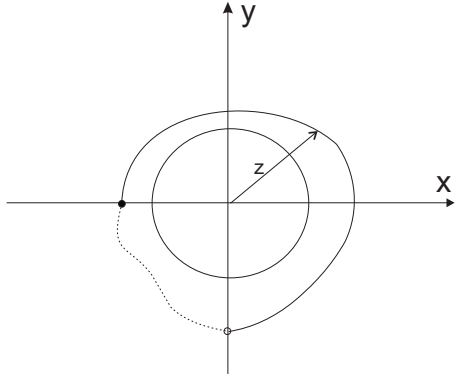}
\end{center}
\caption{A general solution of the two-dimensional problem, i.e., a solution of the functional equation \eqref{ro10}.}
\end{figure}

Let us consider the two-dimensional (unphysical) case, first. The simplest way to solve the problem in two dimensions is to analyze a picture (see, Fig.~2).
The length of the vector $\vec{z}$ in Fig.~2 corresponds to the inverse of the one-way velocity of light in the direction of $\vec{z}$.
Thus, e.g.\ the circle corresponds to a constant (independent of the direction) velocity of light. Since Eq.~\eqref{ro10} is a single equation
with four unknowns, the three unknowns are arbitrary and determine the fourth one. For example, the three coordinates: of the black dot,
of the intersections of the solid curve with positive $y$ and positive $x$ axes respectively are arbitrary, and they determine the coordinate of the
white dot satisfying Eq.~\eqref{ro10}. Analogously, the whole solid curve in Fig.~2 is practically arbitrary (it should only be unique as a continuous
function of the angle) and determines the (dotted) segment with negative coordinates $x$ and $y$.
As a side remark, we observe that changing the angle between the arms of the MM interferometer (usually it is $90^\circ$) changes the dotted segment.
Namely, the angle between the vectors $\vec{z}$ pointing at the black dot and the white dot, respectively, is equal to
the greater angle between the arms of the interferometer.
In principle, the whole curve could be discontinuous in two dotted points. But we can easily avoid this possibility by appropriate
deformation of the primary solid curve.
We could also require the mirror symmetry of the curve. For example, the axis of the symmetry could be interpreted as the direction of the
movement of the laboratory frame, in the spirit of the aether philosophy. A general solution would be determined by an arbitrary solid
curve (see Fig.~3) starting at the black dot (as in Fig.~2) and terminating at the intersection with the line $y=x$.
Mirror reflection with respect to the line $y=x$ reproduces the rest of the curve except the last quarter, which should by constructed
according to \eqref{ro10} as described earlier in this paragraph. One can easily check that, thanks to the mirror symmetry, this time, the whole curve
is automatically continuous provided the primary segment is continuous.

\begin{figure}
\begin{center}
\includegraphics[width=6cm]{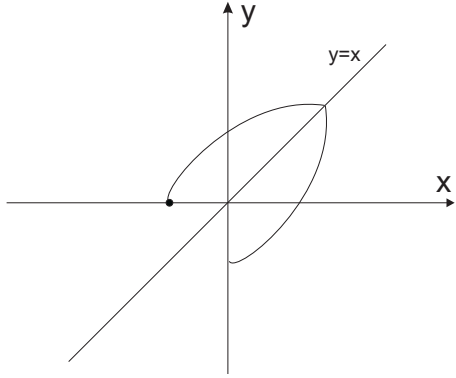}
\end{center}
\caption{A general axially symmetric two-dimensional solution.}
\end{figure}

Up to now we have been considering a(n unphysical) two-dimensional construction. It appears that in three dimensions constraints are a bit
stronger, and the three-dimensional case is qualitatively quite different.
Therefore, we present now an explicit three-dimensional analysis.
We will determine directional dependence of the one-way
velocity of light consistent with MM-type experiments. Thus, roughly, we are interested in (generally) non-constant (continuous) functions on a
sphere ${\cal S}^2$, \mbox{$z(x)\neq{const}$}, defining (the inverse of) the velocity of light in the direction associated to $x\in {\cal S}^2$, i.e.,
if we place the interferometer in the center of the interior of ${\cal S}^2$,
the values of the all four $z$'s ($z_\pm^\perp$, $z_\pm^\shortparallel$) are given by the values of the function $z(x)$
for $x$ belonging to the points of ${\cal S}^2$ corresponding to appropriate axes of the interferometer. The only constraint for the values of $z(x)$ is
given by Eq.~\eqref{ro10}.
For any configuration of the interferometer two $z$'s correspond to the upper hemisphere of ${\cal S}^2$ and the other two to the lower one.
The two upper points uniquely determine the positions of the other two. Therefore, we can confine ourselves to consideration
of the two points on the upper hemisphere. We can project the upper hemisphere onto a two-dimensional disc ${\cal D}^2$. Let us
now define an auxiliary function on ${\cal D}^2$, $\bar{z}(y)$, $y\in {\cal D}^2$, where informally $\bar{z}=z_++z_-$, i.e.\ the value
of $\bar{z}$ is the sum of the opposite $z$'s.
It is easy to show that $\bar{z}$ has to be a constant function on ${\cal D}^2$. It indirectly follows from Eq.~\eqref{ro10} by
virtue of transitivity. Namely, we can connect arbitrary two points B and C on ${\cal D}^2$, and compare the corresponding values
of $\bar{z}$ using an additional auxiliary point A, and next apply Eq.~\eqref{ro10} to the both pairs (see ``kitty'' Fig.~4).
The pair $B$, $C$ does not, in general, corresponds to a position of the interferometer, but the both pairs $B$, $A$
and $C$, $A$, by construction, do. Since $\bar{z}(B)=\bar{z}(A)$ and $\bar{z}(C)=\bar{z}(A)$, then $\bar{z}(B)=\bar{z}(C)$,
and $\bar{z}=const$, i.e.\ $\bar{z}=2c^{-1}$.

\begin{figure}
\begin{center}
\includegraphics[width=5cm]{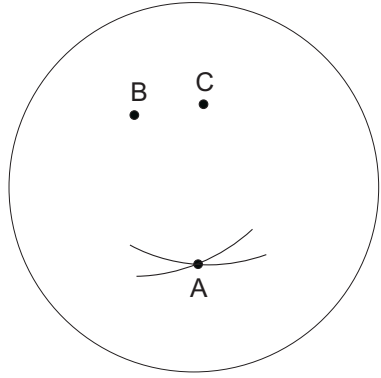}
\end{center}
\caption{The ``angular distance'' between $B$ and $A$, as well as between $C$ and $A$ equals $90^\circ$ because
the both pairs correspond to possible positions of the interferometer.}
\end{figure}

The constancy of $\bar{z}$ is a strong restriction, not having a counterpart in two dimensions. To decipher
this restriction let us now consider an auxiliary function $z_+$ on ${\cal D}^2$ corresponding to one of $z$'s entering
the sum defining $\bar{z}$, say $z_+$, then $z_-=\bar{z}-z_+$. The (continuous) function $z_+$ is almost arbitrary and the
only restriction, of topological nature, is coming from the fact that the sum of the values of $z_+$ on opposite sides of
the boundary of ${\cal D}^2$ should be equal $2c^{-1}$.

This simple informal discussion will be summarized and refined now in more mathematical terms. First of all, ``in the first
approximation'', we can observe that the constant function $\bar{z}$  on ${\cal D}^2$ is actually a constant function on
a two-dimensional real projective surface ${\mathbb R}P^2$. It should seem obvious, because identification of opposite
points of ${\cal S}^2$ provides ${\mathbb R}P^2$ by definition, or in other words, we are interested in functions on a set of rays
rather than on a set of directions. For further convenience, we shift the function (by $2c^{-1}$) down to zero, which is a kind of additive
normalization. What is less obvious, we claim that we now deal with a twisted real linear bundle ${\mathcal B}$ over the
base manifold ${\mathbb R}P^2$. The ``shifted'' $\bar{z}$ denoted as $\tilde{\bar{z}}$ is a zero cross-section of
${\mathcal B}$, whereas the (shifted) function $z_{+}$ denoted as $\tilde{z}_+$ becomes an arbitrary continuous
cross-section of ${\mathcal B}$. The form of (the twist) of the bundle ${\mathcal B}$ follows from the observation that the
values of $\tilde{z}_+$ on opposite sides of the boundary of ${\cal D}^2$ (at the identified points) should be opposite, i.e.,
the non-trivial element of the discrete group ${\mathbb Z}_2$, coming from the corresponding principal bundle, acts in
the fiber $\mathbb R$ (Fig.~5).
To exclude negative velocities of light, and consequently negative times of the travel of light, we can limit the real values
of $\tilde{z}_+$'s to the interval $\left|\tilde{z}_+\right|\le c^{-1}$, and we can speak on the interval
bundle $\bar{\bar{{\mathcal B}}}$ instead of the linear bundle ${\mathcal B}$.

Recapitulating, we could state that all solutions of the problem (angular-dependent one-way velocities of light consistent with
the null-effect of the MM experiment) are parameterized
by cross-sections of the non-trivial bundle $\bar{\bar{{\mathcal B}}}$.
Obviously, non-constant solutions (non-zero cross-sections) do exist. We could also choose an axially symmetric solution
on demand. Still the simplest possibility corresponding to Eq.~\eqref{eq:gwiazdka} and Eq.~\eqref{eq:gwiazdka2} is excluded,
but some mild deformations, e.g.\
expressed in the framework of the Mansouri--Sexl test theory with the parameters $a^2=b^2\left(1-v^2\right)^2$,
$d^2=b^2\left(1-v^2\right)$ \cite{trzy}, are allowable.
The two-dimensional case is even less restrictive than three-dimensional one (richer in a sense) but as unphysical is
less interesting.

\begin{figure}
\begin{center}
\includegraphics[width=6cm]{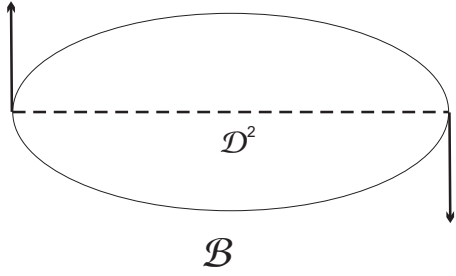}
\end{center}
\caption{The non-trivial element of ${\mathbb Z}_2$ acts at the identified points on opposite sides of the boundary of ${\cal D}^2$.}
\end{figure}

\bigskip
\noindent{\bf Additional remarks}

\noindent
Most of modern versions of the Michelson--Morley experiment is concerned with a particular type of the two-way velocity of light. Namely, light is supposed to travel forth and back along the same way (the trajectory is a contracted loop rather than an arbitrary one), as for example, in each arm of the Michelson--Morley interferometer. We are mainly concerned with the one-way velocity of light: constancy of the particular type of the two-way velocity of light is assumed, and a directional (angular) dependence of the one-way velocity of light is derived.

The ``average'' velocity of light discussed in the beginning of our article is of course a two-way velocity of light. But we wanted to stress that, in contrast to ``standard'' Michelson--Morley experiments, all kinds of loops are admissible.

There seem to be some difficulties with the notion of the one-way velocity of light because no definite synchronization convention has been adopted. Since the only measurements utilized concern the two-way velocity of light a synchronization convention is superfluous. Of course, to measure the one-way velocity of light a synchronization convention is unavoidable. Moreover, the one-way velocity of light (naturally) defines a synchronization convention, and vice versa. Therefore, the analyzed freedom of the one-way velocity of light could have no direct experimental meaning, as for example, gauge in gauge theories. Nevertheless, there are some non-trivial consequences of that freedom, and we mention two of them: a theoretical and an experimental one.

\bigskip
\noindent{Theoretical}

\noindent
Since one can (naturally) relate the one-way velocity of light to a synchronization procedure, there are some restrictions placed upon general clock synchronization convention following from the constraints on directional dependence of the one-way velocity of light \cite{Edwards}. Moreover, one could even speculate about a kind of a possible generalization of the Lorentz transformations to the new synchronization scheme.

\bigskip
\noindent{Experimental}

\noindent
We have assumed constancy only of the particular type of the two-way velocity of light (see, the first paragraph), and an arbitrary average (two-way) velocity of light could, in principle, vary. Therefore, non-standard Michelson--Morley experiments, i.e. those utilizing light travelling along arbitrary, non-contracted loops, would be of potential interest. Negative results of the Michelson--Morley experiments for all kinds of loops would reduce our results to results of \cite{Edwards}. Negative results for ``standard'' (contracted) loops, and positive ones for non-contracted loops would reinforce our general results.

In the three-dimensional case, we compare arbitrary positions of the Michelson--Morley interferometer. Therefore, our analysis seems to be fully three-dimensional, although rather idealized.

Reference \cite{Consoli} has been singled out as an example of the non-orthodox conclusion concerning the Michelson--Morley experiment. In the paper \cite{Cahill} one can find bibliography to several other works of this type.
Non-negative results of the (``standard'') Michelson--Morley experiment would partially invalidate our analysis giving more freedom to the directional dependence of the one-way velocity of light.

\bigskip

This work is supported by the Polish Ministry of Scientific Research and Information Technology under the grant No. PBZ/MIN/008/P03/2003 and by the University of Lodz.

\end{document}